\documentclass[a4paper,twoside]{article}

\usepackage{epsfig}
\usepackage{subfigure}
\usepackage{calc}
\usepackage{amssymb}
\usepackage{amstext}
\usepackage{amsmath}
\usepackage{amsthm}
\usepackage{multicol}
\usepackage{pslatex}
\usepackage{apalike}
\usepackage{algorithm}
\usepackage[noend]{algpseudocode}
\usepackage{caption}
\usepackage{relsize}
\usepackage{SCITEPRESS}     

\newcommand{\twopartdef}[4]
{
	\left\{
		\begin{array}{ll}
			#1 & \mbox{if } #2 \\
			#3 & \mbox{if } #4
		\end{array}
	\right.
}

\begin{document}

\title{Locality-Sensitive Hashing for Efficient Web Application Security Testing}

\author{\authorname{Ilan Ben-Bassat\sup{1} and Erez Rokah\sup{2}}
\affiliation{\sup{1}School of Computer Science, Tel-Aviv University, Tel-Aviv, Israel}
\affiliation{\sup{2}IBM, Herzliya, Israel}
\email{ilanben@post.tau.ac.il, erezrokah@gmail.com}
}


\abstract{Web application security has become a major concern in recent years, as more and more content and services are available online. A useful method for identifying security vulnerabilities is black-box testing, which relies on  an automated crawling of web applications. However, crawling Rich Internet Applications (RIAs) is a very challenging task. One of the key obstacles crawlers face is the state similarity problem: how to determine if two client-side states are equivalent. As current methods do not completely solve this problem, a successful scan of many real-world RIAs is still not possible. We present a novel approach to detect redundant content for security testing purposes. The algorithm applies locality-sensitive hashing using MinHash sketches in order to analyze the Document Object Model (DOM) structure of web pages, and to efficiently estimate similarity between them. Our experimental results show that this approach allows a successful scan of RIAs that cannot be crawled otherwise.}

\onecolumn \maketitle \normalsize \vfill

\section{\uppercase{Introduction}}
\label{intro}

\noindent The information era has turned the Internet into a central part of modern life. Growing amounts of data and services are available today, making web application development an important skill for both enterprises and individuals. The heavier the reliance on web applications, the higher the motivation of attackers to exploit security vulnerabilities. Unfortunately, such vulnerabilities are very common~\cite{gordeychik2010web}, leading to an increasing need to detect and remediate them.

Black-box security scanners address the problem of ensuring secured web applications. They simulate the behavior of a user crawling an application without access to the source code. When discovering new pages and new content, the scanner performs a series of automated security tests, based on a database of known vulnerabilities. This way, the application can be easily and thoroughly analyzed from a security point of view. See~\cite{bau2010state,doupe2010johnny} for surveys on available scanners.

It is clear that a comprehensive security scan requires, among other factors, both {\it high coverage} and {\it efficiency}. If the crawler cannot reach significant parts of the application in a reasonable time, then the security assessment of the application will be incomplete. In particular, a web crawler should refrain from wasting time and memory resources on scanning pages that are similar to previously visited ones. The definition of redundant pages depends on the crawling purpose. For traditional web indexing purposes, different contents imply different pages. However, for security testing purposes, different sets of vulnerabilities imply different pages, regardless of the exact content of the pages.

While the problem of {\it page similarity} (or {\it state similarity}) is a fundamental challenge for all web crawlers~\cite{pomikalek2011removing}, it has become even more significant since the emergence of a new generation of web applications, often called Rich Internet Applications (RIAs).

RIAs make extensive usage in technologies such as AJAX (Asynchronous JavaScript and XML)~\cite{garrett2005ajax}, which enable client-side processing of data that is asynchronously sent to and from the server. Thus, the content of a web page changes dynamically without even reloading the page. New data from the server can be reflected to the user by modifying the DOM~\cite{nicol2001document} of the page through UI events, albeit no change was done to the URL itself. Although such technologies increase the responsiveness of web applications and make them more user-friendly, they also pose a great difficulty on modern web crawlers. While traditional crawlers focused on visiting all possible URLs, current-day crawlers should examine every client state that can be generated via an execution of a series of HTML events.

The exponentially large number of possible states emphasizes the need to efficiently detect similarity between different application states. Failing to solve the state similarity problem results in a poor scan quality. If the algorithm is too strict in deciding whether two states are similar, then too many redundant states would be crawled, yielding long or even endless scans. On the other hand, if the resemblance relation is too lax, the crawler might mistakenly deem new states as near-duplicates of previously-seen states, resulting in an incomplete coverage.

As stated before, every URL of a RIA is actually a dynamic application, in which states are accessed through UI events. Consequently, eliminating duplicate content cannot rely solely on URL analysis, such as in~\cite{bar2009not}. Instead, a common method that black-box scanners use in order to decide whether two pages are similar, is to analyze their DOM structure and properties, without taking the text content into account. This is a useful approach for security oriented crawling, since often the text content of a page is irrelevant to the security assessment. As opposed to the content, the elements and logic that allow user interaction are more relevant for a security analysis. For example, consider a commercial web application with a catalog section, containing thousands of pages. The catalog pages share the same template, yet every one of them is dedicated to a different product. While the variety of products may be of interest for search engines like Google or Yahoo, it is of no great importance to black-box security scanners, as all pages probably share the same set of security vulnerabilities. Therefore, an analysis of the DOM structures of the pages might suggest that they are near-duplicates of each other.

In the past decade, several methods have been proposed to determine similarity between DOMs (see section~\ref{otherwork}). Most of the methods include DOM normalization techniques, possibly followed by applying simple hash functions. While these methods were proven to be successful in scanning several RIAs, they are often too strict, leading to a state explosion problem and very long scans. Moreover, since hash functions used in this context so far have, to the best of our knowledge, no special properties, minor differences between two canonical DOM forms might result in two completely different hash values. Therefore, many complex applications still cannot be crawled in a reasonable time. Approaches involving distance measures, such as~\cite{mesbah2008exposing}, are not scalable and require computing distances between all-pairs of DOMs.

In this paper we present a different approach for the state similarity problem for security oriented crawling. Our approach is based on {\it locality-sensitive hashing} (LSH)~\cite{indyk1998approximate}, and on {\it MinHash sketches} in particular~\cite{broder1997resemblance,broder2000min}. A locality-sensitive hash function satisfies the property that the probability of
collision, or at least close numerical hash values, is much higher for similar objects than for other objects. LSH schemes have already been used in the context of detecting duplicate textual content (see~\cite{pomikalek2011removing} for a survey). Recently, MinHash sketches have also become an efficient solution for bioinformatic challenges, where large amounts of biological sequence data require efficient computational methods. As such, they can be used to detect inexact matches between genomic sequences~\cite{berlin2015assembling,popic2016privacy}. However, LSH techniques have not yet been used by black-box security scanners. The flexibility of MinHash sketches enables detecting duplicate states if two DOMs differ in a small number of DOM elements, regardless of their type. The LSH technique that we use makes the algorithm scalable for large RIAs as well. Combined together, our method makes exploring industrial RIAs feasible. This paper is an extended version of the one appeared in the proceedings of ICISSP 2019 \cite{benbassat2019lsh}.

\section{\uppercase{Related Work}}
\label{otherwork}

\noindent A common approach for an efficient detection of duplicate states in AJAX applications is to apply simple hash functions on the DOM string representation. The authors of \cite{duda2009ajax} and \cite{frey2007indexing} compute hash values according to the structure and content of the state. However, although these methods can remove redundant copies of the same state, they are too strict for the purpose of detecting near-duplicate pages.

CRAWLJAX \cite{mesbah2008exposing} is a crawler for AJAX applications that decides whether two states are similar according to the {\it Levenshtein distance} \cite{levenshtein1966binary} between the DOM trees. Using this method, however, for computing distances between all-pairs of possible states is infeasible in large RIAs. In a later work \cite{roest2010regression}, the state equivalence mechanism of the algorithm is improved by first applying an {\it Oracle Comparator Pipelining} (OCP) before hash values are computed. Each comparator is targeted to strip the DOM string from an irrelevant substring, which might cause meaningless differences between two states, e.g., time stamps, advertisement banners. CRAWLJAX is therefore less strict in comparing application states, but the comparator pipeline requires manual configuration and adjustment for every scan. FEEDEX \cite{fard2013feedback}, which is built on top of CRAWLJAX, uses {\it tree edit distance} \cite{tai1979tree} to compare two DOM trees.

The notion of Jaccard similarity is used in j{\"A}k~\cite{pellegrino2015jak}. In this paper, the authors consider two pages as similar if they share the same normalized URL and their Jaccard index is above a certain threshold. A normalized URL is obtained after stripping the query values and sorting the query parameters lexicographically. The Jaccard similarity is computed between sets of JavaScript event, links and HTML forms that appear in the web pages.

Two techniques to improve any DOM based state equivalence mechanism are presented in \cite{choudhary2012solving}. The first aims to discover unnecessary dynamic content by loading and reloading a page. The second identifies session parameters and ignores differences in requests and responses due to them.

The DOM uniqueness tool described in \cite{ayoub2013document} identifies pages with similar DOM structure by detecting repeating patterns and reducing them to a canonical DOM representation, which is then hashed into a single numerical value. The user can configure the algorithm and determine which HTML tags are included in the canonical representation, and whether to include their text content. This method captures structural changes, such as additions or deletions of rows in a table. It is also not affected by elements shuffling, since it involves sorting the elements in every DOM subtree. However, modifications that are not recognized as part of a structural pattern lead to a false separation between two near-duplicate states.  The method in \cite{moosavi2014indexing} further extends this algorithm by splitting a DOM tree into multiple subtrees, each corresponding to an independent application component, e.g., widgets. The DOM uniqueness algorithm is applied on every component independently, thus avoiding explosion in the number of possible states when the same data is being displayed in different combinations.

The structure of a page is also the key for clustering similar pages in \cite{doupe2012enemy}, in which a model of the web application's state machine is built. The authors model a page using its links (anchors and forms), and store this information in a prefix tree. These trees are vectorized and then stored in another prefix tree, called the Abstract Page Tree (APT). Similar pages are found by analyzing subtrees of the APT.

A different approach for clustering application states into different equivalence clusters appears in software tools that model and test RIAs using execution traces, such as RE-RIA \cite{amalfitano2008reverse}, CrawlRIA \cite{amalfitano2010rich}, and CreRIA \cite{amalfitano2010iterative}. The clustering is done by evaluating several {\it equivalence criteria}, which depend on the DOM set of elements, event listeners and event handlers. Two DOMs are considered equivalent if one set contains the other as a subset. This method has a high memory consumption and computation time.

Research efforts have been made during the years in detecting near-duplicate text, especially in the context of the Web. As the general problem of duplicate content detection was not the focus of this paper, we refer the readers to a detailed survey for more information on the subject \cite{pomikalek2011removing}.

\section{\uppercase{MinHash sketches}}
\label{framework}

\noindent MinHash is an efficient technique to estimate the {\it Jaccard similarity} between two sets. Given two sets, $A$ and $B$, the Jaccard similarity~\cite{jaccard1901distribution} of the sets, $J(A,B)$, is a measure of how similar the sets are:

\begin{equation}
J(A,B) = \frac{|A \cap B|}{|A \cup B|}
\end{equation}

The Jaccard similarity value ranges between 0 (disjoint sets) and 1 (equal sets). However, direct computation of this ratio requires iterating over all the elements of $A$ and $B$. MinHash~\cite{broder1997resemblance} is an efficient method to estimate the Jaccard similarity of two sets, without explicitly constructing their intersection and union. The efficiency of the method comes from reducing every set to a small number of fingerprints.

Let $S$ be a set, and let $h$ be a hash function whose domain includes the elements of $S$. We define $h_{min}(S)$ as the element $a$ of $S$, which is mapped to the minimum value, among all the elements of $S$:

\begin{equation}
h_{min}(S) = arg min_{a \in S}h(a)
\end{equation}

We consider again two sets of elements, $A$ and $B$. One can easily verify that:

\begin{equation}
\label{prob_equal}
Pr[h_{min}(A) = h_{min}(B)] = J(A,B)
\end{equation}

Eq.~\ref{prob_equal} implies that if we define a random variable $R_{A,B}$ as follows:

\begin{equation}
\label{rv}
R_{A,B} = \twopartdef { 1\,\, } {\mbox h_{min}(A) = h_{min}(B) }{ 0\,\, } {\mbox{otherwise}}
\end{equation}

\noindent then $R_{A,B}$ is an indicator variable which satisfies $E[R_{A,B}] = J(A,B)$. However, $R_{A,B}$ is either 0 or 1, so it is not useful as an estimator of the Jaccard similarity of $A$ and $B$. As it is done in many estimation techniques, it is possible to reduce the variance of $R_{A,B}$ by using multiple hash functions and computing the ratio of the hash functions, for which the minimum element is the same. In other words, assume now a set of $\ell$ hash functions, $h^{1}, \dots , h^{\ell}$. For each $1 \leq i \leq \ell$ we define $R_{A,B}^{(i)}$ as in Eq.~\ref{rv}, replacing $h$ with $h^{i}$. The improved estimator for the Jaccard similarity of $A$ and $B$, $T(A,B)$, is now given by:

\begin{equation}
\label{estimator}
T(A,B) = \frac{\sum_{i=1}^{\ell}{R_{A,B}^{(i)}}}{\ell}
\end{equation}

By Chernoff bound, it can be shown that the expected error is $O(\frac{1}{\sqrt{\ell}})$.

The compressed representation $\langle h^{1}_{min}(S), \dots, h^{\ell}_{min}(S) \rangle$ of a set $S$ is defined as its MinHash sketch. Since $\ell$ is significantly smaller than the size of $S$, the estimation of the Jaccard similarity is efficient in both time and memory.

\section{\uppercase{Method}}
\label{method}

\noindent In this section we present the complete algorithm for detecting duplicate content during a black-box security scan. The first step is transforming a web page, given as an HTTP response string, into a set of {\it shingles}, or {\it $k$-mers}. In the second step, the algorithm uses MinHash sketches in order to efficiently compute similarity between pages. However, the method described in section~\ref{framework} still requires that we compute the similarity between all possible pairs of states. Instead, we use an efficient LSH approach that focuses only on pairs that are potentially similar. We outline the complete algorithm at the end of this section.

\subsection{Set Representation of Web Pages}
\label{page2set}

\noindent As stated in section~\ref{intro}, the text content of a web page is usually irrelevant for the state similarity problem in the context of security scans. Therefore, we rely on the DOM representation of the page. The algorithm extracts the DOM tree from the HTTP response, and serializes it to a string. The relevant information for the state similarity task includes the DOM elements and their structure, the events and the event handlers that are associated with the elements, as well as some of their attributes. Yet, for simplicity reasons, in this paper we only consider the names of the elements and the structure of the DOM. 

\begin{figure*}
\includegraphics[width=\textwidth,height=7cm]{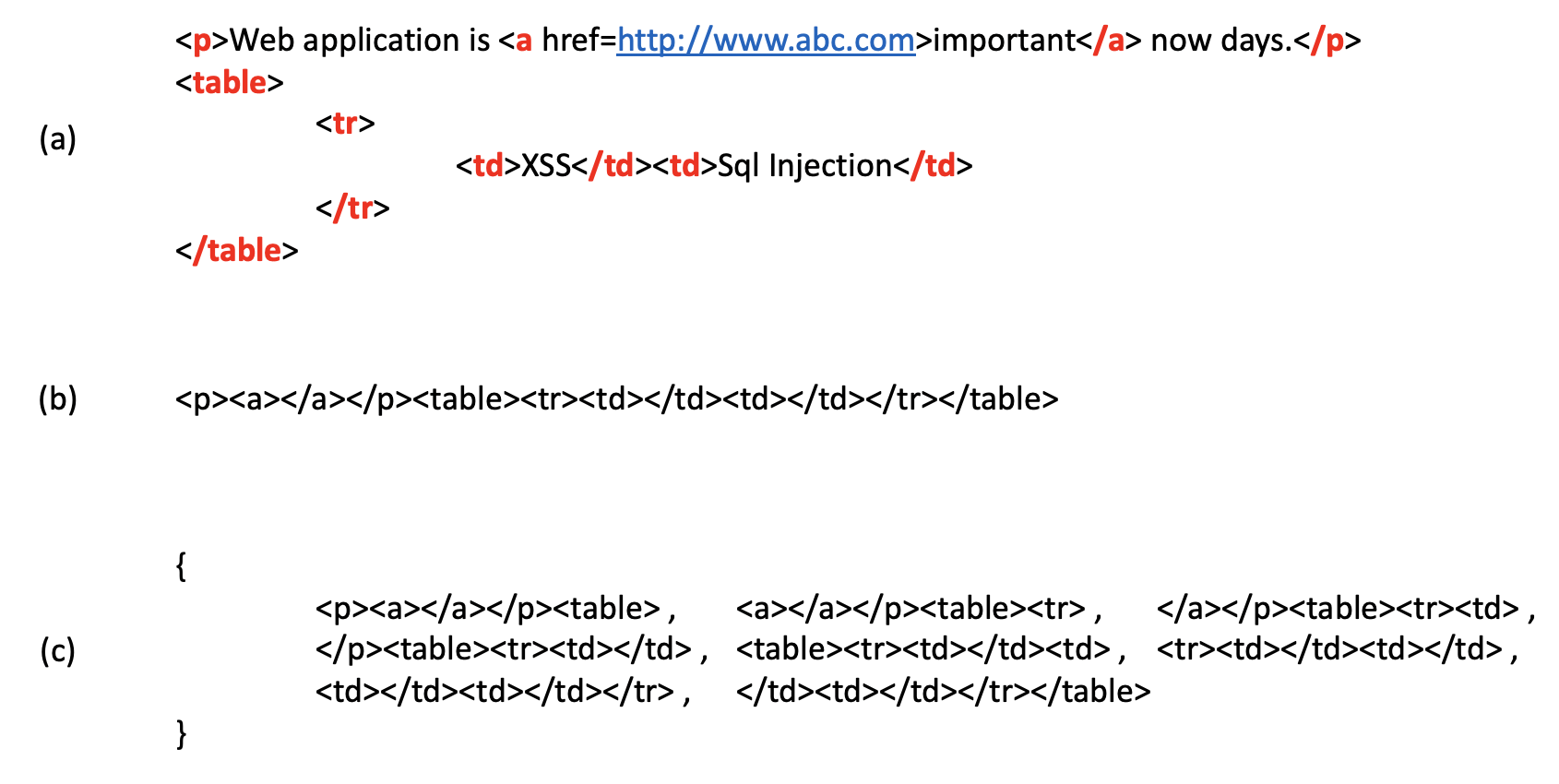}
\caption{Constructing a set representation of an HTTP response. (a) HTTP response snippet. The elements of the DOM tree are marked in red. (b) Serialization of the corresponding DOM tree to a string. (c) The set of all 5-mers of consecutive DOM elements. This is the set representation of the response for $k=5$.}
\label{page-to-set}
\end{figure*}

Since we are interested in using MinHash, there is a need to transform the string representation of the DOM into a set of elements. For the sake of this purpose, we use the notion of shingles, or $k$-mers, which are sequences of any $k$ consecutive words. In this case, since the DOM elements are the building blocks of the DOM tree, we consider every element as a word. The algorithm can filter out part of the DOM elements, if they are marked as irrelevant. Figure~\ref{page-to-set} illustrates the process of constructing the set representation of an HTTP response.

A key factor in the performance of the algorithm is the choice of the value of $k$. If $k$ is too small, then many $k$-mers will appear in all web pages, and most of the pages will be similar to each other.  Alternatively, as the value of $k$ becomes too high, the comparison of states is too strict. So, $k$ should have a large enough value, such that the probability of different states sharing the same $k$-mer, is low.

\subsection{Efficient LSH for MinHash Sketches}
\label{lsh4minhash}

\noindent To accelerate the process of detecting duplicate content, we use a hash table indexing approach, which rearranges the sketches in a way that similar states are more likely to be stored closely in our data structure. Dissimilar states, on the other hand, are not stored together, or are rarely stored in proximity. Such an approach is also used in other domains, e.g., genome assembly~\cite{berlin2015assembling}.

The algorithm constructs a data structure of $\ell$ hash tables, each corresponding to a different hash function. In the $i$-th hash table, we map every hash value $v$, which was computed by the $i$-th hash function, to a set of DOM IDs, which correspond to DOMs that are hashed to this value. So, if we denote the set of all DOMs by $\mathcal{P}$, and for every set-representation $P \in \mathcal{P}$ we denote its ID by $ID(P)$, then the $i$-th table is a mapping of the following form:

\begin{equation}
\label{mapping}
v \mapsto \left\{ ID(P) \mid P \in \mathcal{P} \land h^{i}_{min}(P) = v \right\}
\end{equation}

Using Eq.~\ref{prob_equal}, we can see that the higher the Jaccard similarity of two DOMs is, the higher the probability of their IDs being in the same set (or {\it bucket}).
Given two DOM representations, $P$  and $P'$, we get by Eq.~\ref{estimator} that an estimator for the Jaccard similarity $J(P,P')$ can be computed according to the number of hash tables, for which $ID(P)$ and $ID(P')$ are in the same bucket. This follows directly from the fact that for every $1 \leq i \leq \ell$, if the $i$-th hash table contains both $ID(P)$ and $ID(P')$ in the same bucket, then it holds that $h^{i}_{min}(P) = h^{i}_{min}(P')$. So, we can derive an estimator, $T^{*}$, for this particular LSH scheme, which is based on Eq.~\ref{estimator}. Let us denote the $i$-th hash table by $g^{i}$, $1 \leq i \leq \ell$, and let us mark the bucket which is mapped to hash value $v$ in the $i$-th table, by $g^{i}(v)$. We then have:

\begin{equation}
\label{lsh-estimator}
T^{*}(P,P') = \frac{\left\vert\big\{i \mid \exists{v}. {ID(P), ID(P')} \subseteq g^{i} (v)\big\}\right\vert}{\ell}
\end{equation}

Following Eq. \ref{lsh-estimator}, there is no need to compare every DOM to {\it all} other DOMs. For every pair of DOM representations, which share no common bucket, the estimated value of their Jaccard similarity is 0. 

\subsection{The Complete Algorithm}
\label{algo}

\noindent We now combine the shingling process and the LSH technique, and describe the complete algorithm for detecting duplicate content, during a black-box security scan. For simplicity, we assume a given set of application states, $\mathcal{S}$. With this assumption in mind, we can discard the part of parsing web pages and extracting links and JavaScript actions. A pseudo-code of the algorithm is given in Algorithm~\ref{crawling}.

In an initialization phase (lines 1-3), the code generates $\ell$ hash functions, out of a family $\mathcal{H}$ of hash functions. For each hash function, $h^{i}$, a corresponding hash table, $g^{i}$, is allocated in an array of hash tables. The rest of the code (lines 4-20) describes the crawling process and the creation of an index of non-duplicate pages.

For each application state, $s$, the algorithm creates its DOM set representation, $P$, using the method {\it Shingle($s$,$k$)}. The method extracts the DOM tree of the web page of state $s$, filters out all text and irrelevant elements, and converts the list of DOM elements into a set of overlapping $k$-consecutive elements. We omit the description of this method from the pseudo-code.

lines 6-13 analyze the DOM set representation, $P$. A MinHash sketch of the set of shingles is computed by evaluating all hash functions. While doing so, we maintain a count of how many times every DOM ID shares the same bucket with the currently analyzed DOM, $P$. This is done using a mapping, $f$, which is implemented as a hash table as well, with a constant-time insertion and lookup (on average). The highest estimated Jaccard similarity score is then found (line 14). If this score is lower than the minimum threshold, $\tau$, then application state $s$ is considered to have new content for the purpose of the scan. In such a case, it is  added to the index, and the data set of hash values is updated by adding the MinHash sketch of the new state. Otherwise, it is discarded as duplicate content.


\begin{algorithm}
\small
\caption{Removing duplicate web application states.}
\label{crawling}
\begin{algorithmic}[1]

\Require{$\mathcal{S}$: set of web application states$\newline\qquad \mathcal{H}$: family of hash functions $\newline\qquad k$: shingle size $\newline\qquad \ell$: MinHash sketch size $\newline\qquad \tau$: Jaccard similarity threshold}

\Ensure{set of states with no near-duplicates}

\State  $\langle h^{1}, \dots, h^{\ell} \rangle \leftarrow$  sample $\ell$ functions from $\mathcal{H}$
\State $[ g^{1}, \dots, g^{\ell}]$ $\leftarrow$ array of $\ell$ hash tables
\State {\it index} $\leftarrow$ $\emptyset$

\For {{\it $s$} $\in$ {\it $\mathcal{S}$}}
	\State $P \leftarrow$ {\it Shingle($s,k$)}
	\State $f \leftarrow$ mapping of type $\mathbb{N} \rightarrow \mathbb{N}^{+}$

	\For {{\it $i$} in range $1, \dots , \ell$}
		\State {\it $v$}  $\leftarrow$ $h^{i}_{min}(P)$
		\For {{\it docId} in {\it $g^{i}$}({\it $v$})}
			\If {{\it docId} $\in f$}
				\State $f(docId) \leftarrow f({\it docId})$ + 1
			\Else
				\State $f(docId) \leftarrow$ 1
			\EndIf
		\EndFor
	\EndFor

	\State {\it score} $\leftarrow$ $\max_{j \in f}{f(j)}$

	\If{{\it score} $ < \tau\ell$}

		\For {{\it $i$} in range $1, \dots , \ell$}
			\State {\it $v$}  $\leftarrow$ $h^{i}_{min}(P)$
			\If {{\it v} $\notin g^{i}$}
				{\it $g^{i}$}({\it $v$}) $\leftarrow \emptyset$
			\EndIf
			\State {\it $g^{i}$}({\it $v$}) $\leftarrow$ {\it $g^{i}$}({\it $v$})  $\cup$ $\left\{ ID(s) \right\}$			
		\EndFor
		\State {\it index} $\leftarrow$ {\it index} $\cup$ $\left\{ s \right\}$
	\EndIf
\EndFor

\State \textbf{return} {\it index}

\end{algorithmic}
\end{algorithm}

The state equivalence mechanism described here is not an equivalence relation, since it is not transitive. This fact implies that the order in which we analyze the states can influence the number of unique states found. For example, for every $\tau$ and every odd integer $m$, we can construct a series of application states $s_{1}, \dots, s_{m}$, such that $J(s_{i}, s_{i+1}) \geq \tau$ for every $1 \leq i \leq m-1$, but $J(s_{i}, s_{j}) < \tau$ for every $1 \leq i,j \leq m$ such that $\left\vert i-j \right\vert \geq 2$. Algorithm \ref{crawling} outputs either $\lfloor \frac{m}{2} \rfloor$ or $\lceil \frac{m}{2} \rceil$ unique states, depending on the scan order. Although theoretically possible, we argue that web applications rarely exhibit such a problematic scenario.

\subsection{MinHash Generalization Properties}
\label{generalization}

\noindent It is clear that the MinHash algorithm generalizes na\"ive methods that directly apply hash functions on the entire string representation of the DOM, such as~\cite{duda2009ajax} and~\cite{frey2007indexing}. By generalization we mean that for any given pair of states, $s_1$ and $s_2$, and any given hash function $h$, if $h(s_1) = h(s_2)$, then applying our LSH scheme also yields an equality between the states. Therefore, any pair of pages that are considered the same state by a na\"ive hashing algorithm, will also be treated as such by the method we propose.

More interesting is the fact that our algorithm is also a generalization of a more complex method. The algorithm in~\cite{ayoub2013document} identifies repeating patterns that should be ignored when detecting duplicate content. Denote by $d_1$ and $d_2$ two DOM strings that differ in the number of times that a repeating pattern occurs in them. More precisely, let $d_1 = ARRB$ and $d_2 = ARRR..RRRB$ be two DOM strings, where $A$, $B$, and $R$ are substrings of DOM elements. Let us assume that the method in~\cite{ayoub2013document} identifies the repeating patterns and obtains the same canonical form for both $d_1$ and $d_2$, which is $ARB$. This way, $d_1$ and $d_2$ are identified as duplicates by~\cite{ayoub2013document}. It is easy to see then that if $k < 2\left|R\right|$, where the length of a DOM substring is defined as the number of DOM elements in it, then the MinHash approach will also mark these two DOM strings as duplicates, since there is no $k$-mer that is included in $d_2$ and not in $d_1$, and vice versa. This proof does not hold for the case of $d_1 = ARB$. However, our approach will also mark $d_1$ and $d_2$ in this case as near-duplicate content with high probability if $k$ is relatively small comparing to the lengths of $A$, $B$, and $R$.

\section{\uppercase{Performance evaluation}}
\label{results}

\noindent In this section we present the evaluation process of the LSH based approach for detecting similar states. We report the results of this method when applied to real-world applications, and compare it to four other state equivalence heuristics.

\subsection{Experimental Setup}

\noindent We implemented a prototype of the LSH mechanism for MinHash sketches as part of IBM \textsuperscript{\scriptsize{\textregistered}} Security AppScan  \textsuperscript{\scriptsize{\textregistered}} tool \cite{IBMAS}. AppScan uses a different notion for crawling than most other scanners. It is not a request-based crawler, but rather an action-based one. The crawler utilizes a browser to perform actions, instead of manipulating HTTP requests and directly executing code via a JavaScript engine. While processing a new page, all possible actions, e.g., clicking on a link, submitting a form, inputting a text, are extracted and added to a queue of unexplored actions. Every action is executed after replaying the sequence of actions that resulted in the state from which it was originated. The crawling strategy is a mixture of BFS and DFS. As a result, the crawler can go deeper into the application, while still avoid focusing on only one part of it.

The MinHash algorithm can be combined with any crawling mechanism. The algorithm marks redundant pages, and their content is ignored. Offline detection of duplicate content is not feasible for modern, complex applications with an enormous amount of pages, since one cannot first construct the entire set of possible pages. In fact, this set can be infinite, as content might be dynamically generated.

We performed security scans on seven real-world web applications and compared the efficiency of our method with other DOM state equivalence algorithms. The first three applications are simple RIAs, which can be manually analyzed in order to obtain a true set of all possible application states. We were given access to two IBM-internal applications for managing security scans on the cloud: {\it Ops Lead} and {\it User Site}. As a third simple RIA we chose a public web-based file manager called {\it elFinder} \cite{elfinder}. We chose applications that are used in practice, without any limitations on the web technologies they are implemented with. 

In order to assess the performance of a state equivalence algorithm, it must also be tested on complex applications with a significant number of near-duplicate pages. Otherwise, inefficient mechanisms to detect similar states might be considered successful in crawling web applications. Therefore, we also conducted scans on four complex online applications: {\it Facebook}, the famous social networking website \cite{FCB}; {\it Fandango}, a movie ticketing application \cite{fandango}; {\it GitHub}, a leading development platform for version control and collaboration \cite{github}; and {\it Netflix}, a known service for watching TV episodes and movies \cite{netflix}, in which we crawled only the {\it media} subdomain. These applications were chosen due to their high complexity and extensive usage of modern web technologies, which pose a great challenge to web application security scanners. In addition, they contain a considerable amount of near-duplicate content, which is not always the case when it comes to offline versions of real-world web applications. We analyzed the results of scanning the four complex applications without comparing them to a manually-constructed list of application states. 

Since a full process of crawling a complex RIA can take several hours or even days, a time limit was set for every scan. Such a limit also enables to test how fast the crawler can find new content, which is an important aspect of a black-box security scanner. 

We report the number of non-redundant states found in the scans, along with their duration times. However, the number of discovered states does not necessarily reflect the quality of the scan: a state equivalence algorithm might consider two different states as one (false {\it merge}), or treat two views of the same state as two different states (false {\it split}).  Furthermore, the scan may not even reach all possible application states. In order to give a measure of how many false splits occur during a scan, we compute the scan {\it efficiency}. The efficiency of a scan is defined as the fraction of the scan results which contains new content. In other words, this is the ratio between the real number of unique states found and the number of unique states reported during the scan. A too-strict state equivalence relation implies more duplicate content and a lower scan efficiency. The {\it coverage} of the scan is defined as the ratio between the number of truly unique states found and the real number of unique states that exist in the application. If the relation is too lax, then too many false merges occur, leading to a lower scan coverage. Inefficient scans can have low coverage as well, if the scan time is consumed almost entirely in exploring duplicate content. As the scan coverage computation requires knowing the entire set of application states, we computed it only for the three simple applications. The scan efficiency is reported for all scanned applications. 

This paper suggests a new approach for the state similarity problem. Hence we chose evaluation criteria that are directly related to how well the scanner explored the tested applications. We do not assess the quality of the explore through indirect metrics such as the number of security issues found during the scans. In addition, during the test phase AppScan sends thousands of custom test requests that were created during the explore stage. Such amount of requests could overload the application server or even affect its state, and we clearly could not do that for applications like Facebook or GitHub. For these reasons we do not report how many security vulnerabilities were detected in each scan. Another metric that is not applicable for online applications is the code coverage, i.e., the number of lines of code executed by the scanner. This metric cannot be computed as we only have limited access to the code of these complex online applications. However, it is clear that scans that are more efficient and have a higher coverage rate can detect more security vulnerabilities. 

\begin{table*}[h]
\caption{Results of security scans on three simple real-world web applications using five different strategies for DOM state equivalence: j{\"A}k (JK), CRAWLJAX (CJ), Simple Hash (SH), DOM Uniqueness (DU), and MinHash (MH). The first row shows the real number of unique states in the application, whereas the second row contains the number of non-redundant states reported in each scan. The next two lines provide the coverage and efficiency rates of the scans. In the last row, a runtime of thirty minutes corresponds to reaching the time limit.}
{\footnotesize
\begin{center}
\begin{tabular}{|c|c|c|c|c|c|c|c|c|c|c|c|c|c|c|c|}
\cline{1-16}
&  \multicolumn{5}{ c| }{elFinder} & \multicolumn{5}{ c| }{User Site}&  \multicolumn{5}{ c| }{Ops Lead} \\ \cline{2-16}
 & JK & CJ &  SH & DU & \textbf{MH} & JK & CJ &  SH & DU & \textbf{MH} & JK & CJ &  SH & DU & \textbf{MH} \\ \hline\hline
\mbox{App states}&  \multicolumn{5}{ c| }{7} & \multicolumn{5}{ c| }{18}&  \multicolumn{5}{ c| }{27} \\ \hline
\mbox{Scan states} & 7 & 12 & 38 & 34  & \textbf{7} & 18 & 20 &  31 &  21 & \textbf{18} & 31 & 33 & 42  & 41 & \textbf{25} \\ \hline
\mbox{Coverage \text{\scriptsize{(\%)}}} & 100 & 100 & 100 &  100 & \textbf{100} & 100 & 100 & 100  & 100 & \textbf{94} & 100 & 100 & 93  & 89 & \textbf{89} \\ \hline
\mbox{Efficiency \text{\scriptsize{(\%)}}} & 100 & 58 & 18 &  21 & \textbf{100} & 100 & 90 & 58  & 86 & \textbf{94} & 87 & 82 & 60  & 59 & \textbf{96} \\ \hline
\mbox{Time \text{\scriptsize{(m)}}}  & 8 & 7 & 30 &  30 & \textbf{10} & 2 & 2 & 4  & 3 & \textbf{2} & 9 & 10 & 30 & 23 & \textbf{12} \\ \hline
\end{tabular}
\end{center}
}

\label{performance-simple}
\end{table*}

Our proposed method, {\it MinHash}, was compared to four other approaches: two hash-based algorithms, and two additional security scanners that apply non hash-based techniques to detect similar states. 
The {\it Simple Hash} strategy hashes every page according to the list of DOM elements it contains (order preserved). The second hash-based method, the {\it DOM Uniqueness} technique~\cite{ayoub2013document}, identifies similar states by reducing repeating patterns in the DOM structure and then applying a simple hash function on the reduced DOMs. As a third approach we used j{\"A}k~\cite{pellegrino2015jak}, which solves the state similarity algorithm by computing the Jaccard similarity between pages with the same normalized URL. Another non hash-based approach was evaluated by using CRAWLJAX~\cite{mesbah2008exposing}. The crawler component of the latter tool uses the Levenshtein edit distance to compute similarity between pages. We used the default configuration of these scanners. Section~\ref{otherwork} provides more details on the approaches we compared our tool with.

We scanned the tested applications three times, each time using a different DOM state equivalence strategy. All scans were limited to a running time of 30 minutes. In the {\it MinHash} implementation we set the value of $k$ to 12, the number of hash functions, $\ell$, to 200, and the minimum similarity threshold, $\tau$, was set to 0.85. The values of $k$ and $\tau$ were determined by performing a grid search and choosing the values which gave optimal results on two of the tested applications. The number of hash functions, $\ell$, was computed using Chernoff bound, so the expected error in estimating the Jaccard similarity of two pages is at most 7\%.

\subsection{Results for Simple RIAs}

\noindent Table \ref{performance-simple} lists the results of the experimental scans we conducted on three simple RIAs. The results show that using MinHash as the DOM state equivalence scheme yields fast and efficient security scans, with a negligible impact on the coverage, if any. Scanning simple applications with j{\"A}k also produces high-quality results. The rest of the approaches were less efficient, overestimating the number of unique states. Some of the scans involving the other hash-based techniques were even terminated due to scan time limit. However, since the tested applications were very limited in their size, these approaches usually had high coverage rates.

\subsubsection{MinHash Results}
\begin{figure}
\centering
\includegraphics[height=5.4cm,keepaspectratio]
{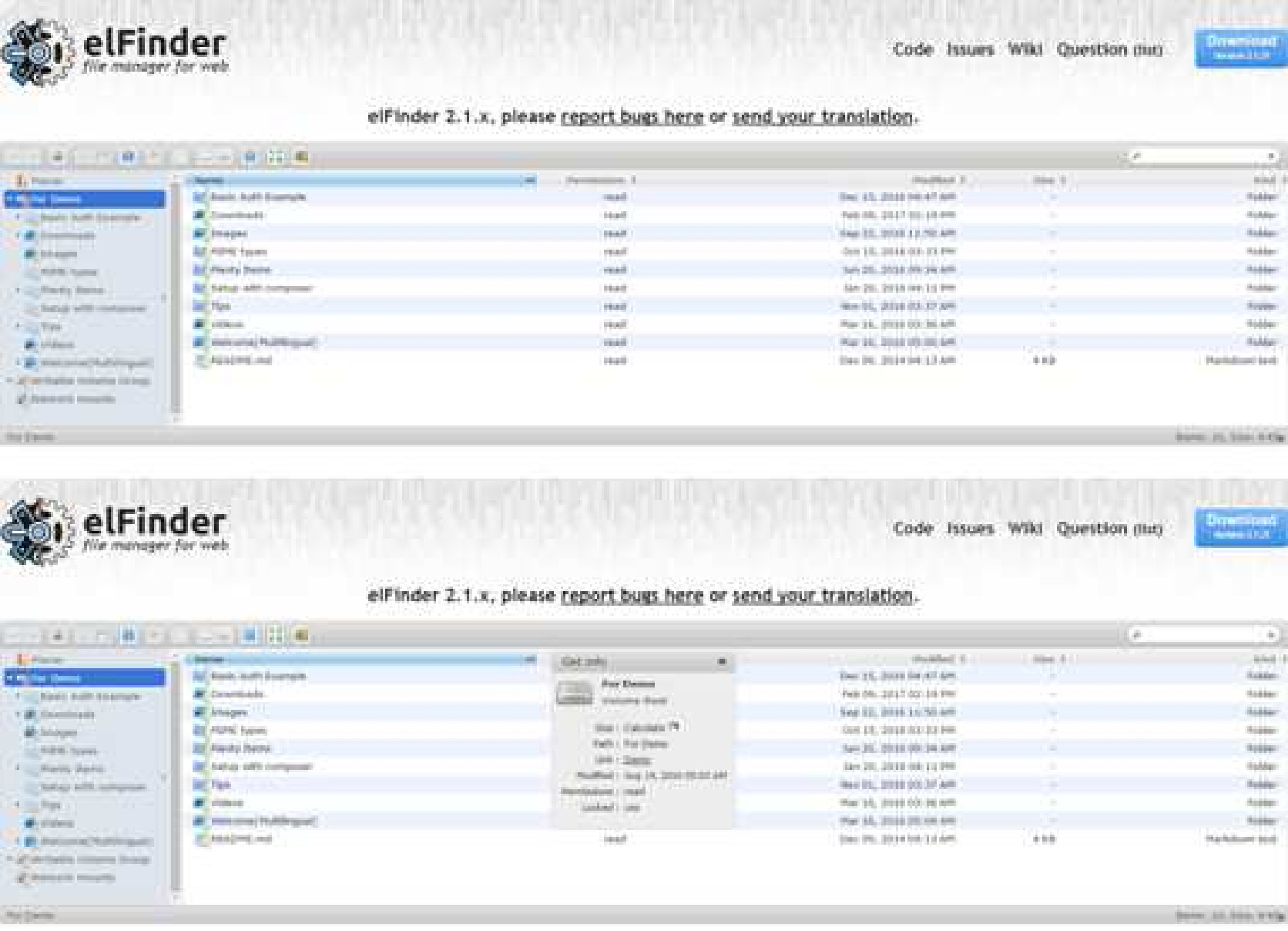}
\caption{Successful merge of two views with the same functionality using the MinHash algorithm. (a) Upper part: a list view of a directory. (b) Lower part: the same view with a textual popup.}
\label{DU-failure}
\end{figure}

\noindent The MinHash scans were very fast and produced concise application models. They quickly revealed the structure of the applications, leading to efficient scans also in terms of memory requirements. The proposed algorithm managed to overcome multiple types of changes between pages that belong to the same state. This way, pages with additional charts, banners, or any other random DOM element, were usually considered as duplicates (see Figure \ref{DU-failure}). Similar pages that differ in the order of their components were also correctly identified as duplicate content, since the LSH scheme is almost indifferent to a component reordering. This property holds since a component reordering introduces new $k$-mers only when new combinations of consecutive components are created. As the number of reordered components is usually significantly smaller than their size, the probability of a change in the MinHash sketch is low. In addition, multiple views of the same state that differ in a repeating pattern were also detected. This is in correspondence with section \ref{generalization}, which shows that our method is a generalization of the DOM Uniqueness approach.  

There were, however, cases where different states were considered the same, and this led to some content being missed. The missing states were very similar to states that had already been reported in the model. See section \ref{conclusions} for a discussion on this matter.

\begin{table*}[h]
\caption{Results of security scans on four complex real-world web applications using five different strategies for DOM state equivalence. The first column for every application contains the number of non-redundant states reported in each scan. The second column provides the efficiency rate (\%). Scans that did not reach the time limit and crashed due to memory or other problems - are marked with asterisk. }
{ 
\begin{center}
\begin{tabular}{|c|c|c||c|c||c|c||c|c|}
\cline{1-9}
& \multicolumn{2}{ c|| }{Facebook} &  \multicolumn{2}{ c|| }{Fandango} & \multicolumn{2}{ c|| }{GitHub} & \multicolumn{2}{ c| }{Netflix} \\ \cline{2-9}
&  States & Efficiency & States &  Efficiency & States & Efficiency & States & Efficiency \\ \hline\hline
\mbox{j{\"A}k} & 1\textsuperscript{*} & 100 & 54 & 37 & 26  & 34 & 47  & 32 \\ \hline
\mbox{CRAWLJAX} & 68 & 38 & 32 & 50 & 184 & 30 & 40 & 40 \\ \hline
\mbox{Simple Hash} & 694\textsuperscript{*} & 2 & 493 & 4  & 306 & 22 & 206 & 7 \\ \hline
\mbox{DOM Uniq.} & 692\textsuperscript{*} & 3 & 468 & 5 & 266 & 30 & 133 & 17 \\ \hline
\textbf{\mbox{MinHash}} & \textbf{200} & \textbf{83} & \textbf{34} & \textbf{82} & \textbf{108} & \textbf{81} & \textbf{27} & \textbf{78} \\ \hline
\end{tabular}
\end{center}
}

\label{performance-complex}
\end{table*}

\subsubsection{Results of Other Hash-Based Approaches}
\noindent The scans produced by the other hash-based methods had high coverage rate. 
However, they were sometimes very long, and the number of reported states was very high (up to a factor of five compared to the correct number of states). Some were even terminated due to time limit. The stringency of the equivalence relations led to inefficient results and to unclear application models. Such models will also result in a longer test phase, as the black-box scan is about to send hundreds and thousands of redundant custom test requests.

As expected, the na\"ive approach of simple hashing was very inefficient.
The DOM Uniqueness method had better results in two out of the three tested applications. However, relevant scans were still long and inefficient (as low as 21\% efficiency rate). Although the algorithm in \cite{ayoub2013document} detects similar pages by reducing repeating patterns in the DOM, it cannot detect similar pages that differ in other parts of the DOM. Consider, for example, two similar pages where only one of them contains a textual popup. Such a pair of pages was mistakenly deemed as two different states (see Figure \ref{DU-failure}). A similar case occurred when the repeating pattern was not exact, e.g., a table in which every row had a different DOM structure).


\subsubsection{Results of Non Hash-Based Tools}
\noindent The scans obtained using j{\"A}k and CRAWLJAX were also very fast and with perfect coverage. However, they were not always as efficient as the MinHash-based scans. The relatively small number of pages and states in the tested applications enabled short running times, although these non hash-based methods perform a high number of comparisons between states. Among the two methods, j{\"A}k was more efficient than CRAWLJAX.

\subsection{Results for Complex RIAs}

\noindent The results shown in Table \ref{performance-complex} emphasize that an accurate and efficient state equivalence algorithm is crucial for a successful scan of complex applications. The MinHash algorithm enabled scanning these applications, while the na\"ive approach and the DOM uniqueness algorithm completely failed in doing so in some cases. CRAWLJAX and j{\"A}k had reasonable results; however, their efficiency rates were not high.

\subsubsection{MinHash Results}
\begin{figure*}
\centering
\includegraphics[width=\textwidth,height=6cm]{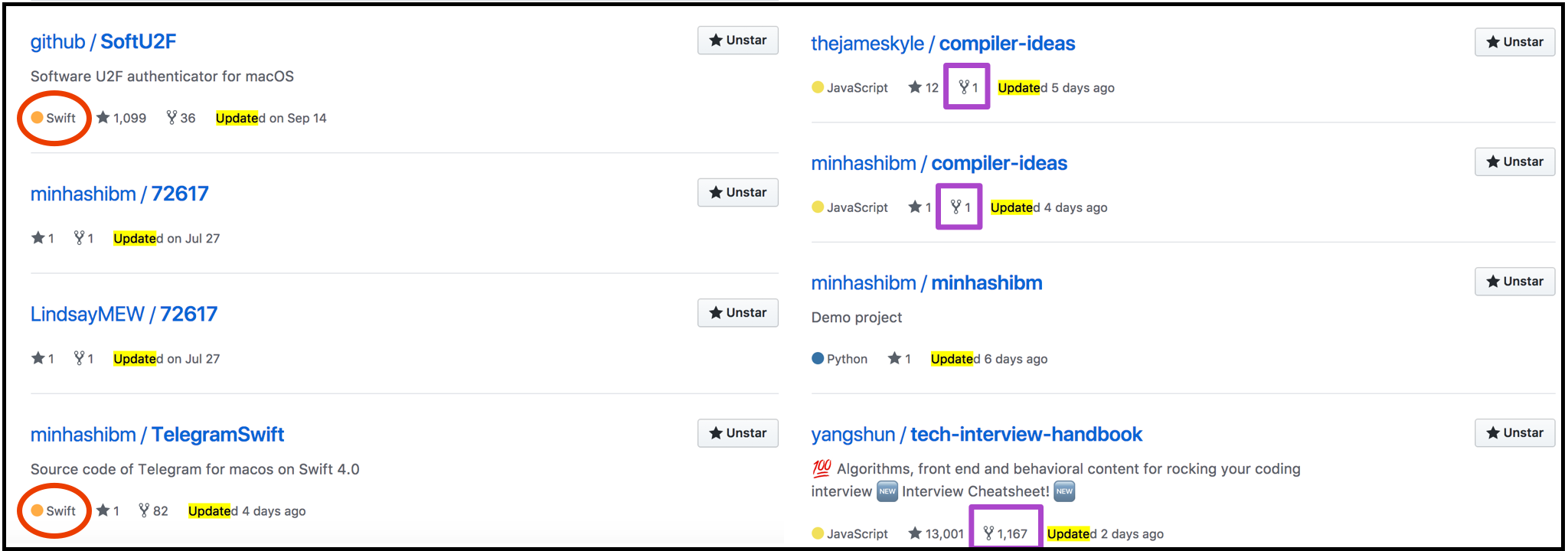}
\caption{Successful merge of two different states using the MinHash algorithm. Both sides of the figure show lists of starred (tracked) GitHub repositories. (a) The left list contains information on the repository programming language for the first and last items only. (b) The right list does not contain information on number of repository copies for the third item. Although the items of the lists are not identical in their structure, the MinHash algorithm can detect that both lists belong to the same application state. On the other hand, the DOM Uniqueness algorithm fails to merge these two states, since the lists do not have the same constant repeating pattern. The DOM structure of every item in every list might be different than the structures of its neighboring items, and thus the items cannot be reduced to a single repeating pattern.}
\label{GitHub-good-merge}
\end{figure*}

\begin{figure}
\centering
\includegraphics[height=7.2cm,keepaspectratio]
{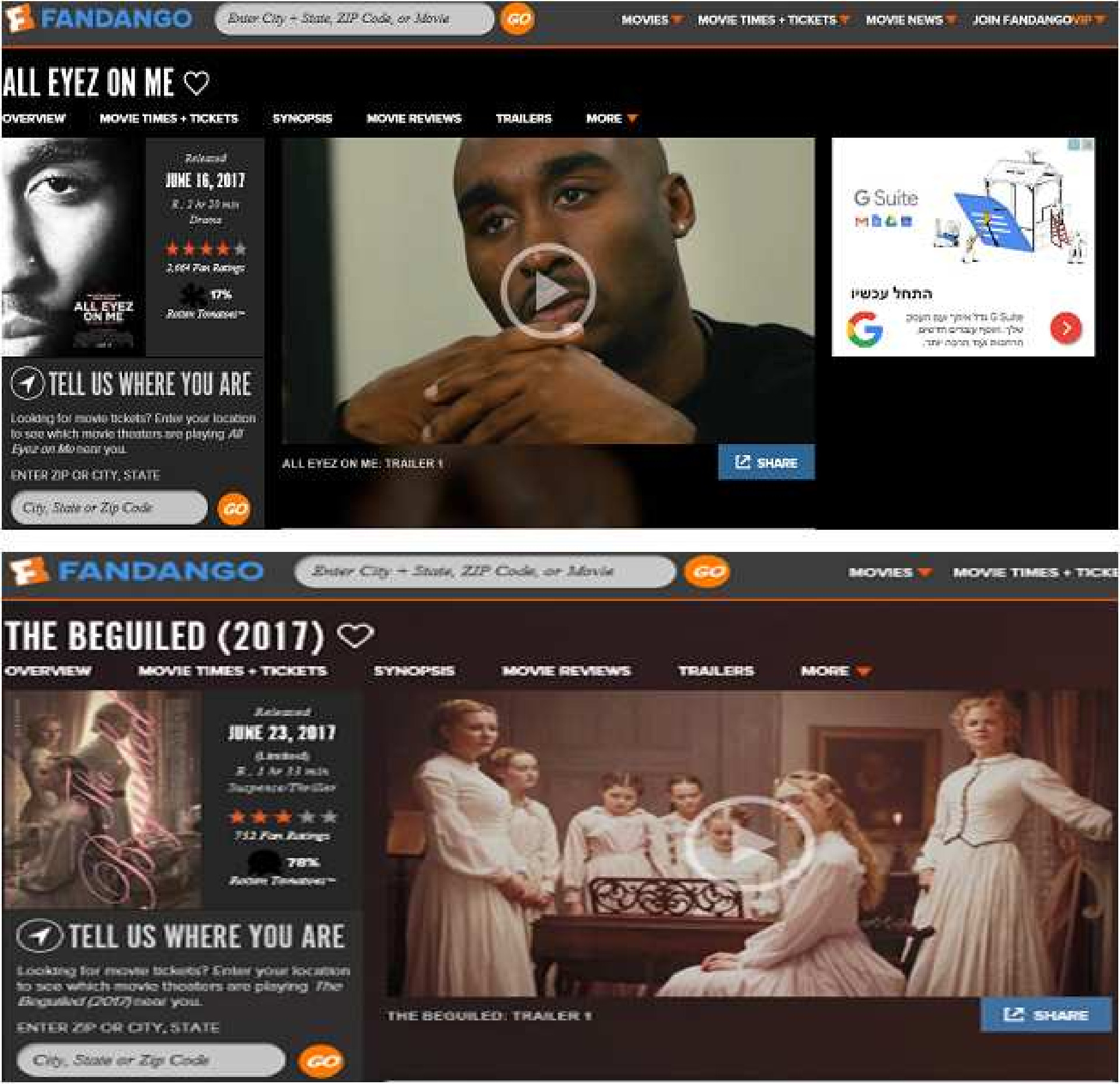}
\caption{Successful merge of two different states using MinHash sketches. (a) Upper part: movie overview page (b) Lower part: an overview page of a different movie. }
\label{FANDANGO-good-merge}
\end{figure}

\noindent The MinHash-based scans efficiently reported dozens of application states, with approximately 80\% of the states being truly unique. In addition, the number of false merges was very low.  The algorithm successfully identified different pages that belong to the same application state. For example, pages that describe different past activities of Facebook users were considered as duplicate content. This is, indeed, correct, as these pages contain the same set of possible user interactions and the same set of possible security vulnerabilities. A typical case in which the MinHash algorithm outperforms other methods can also be found in the projects view of GitHub, as can be seen in Figure \ref{GitHub-good-merge}. A GitHub user has a list of starred (tracked) projects. Two GitHub users may have completely different lists, but these two views are equivalent, security-wise. Figure \ref{FANDANGO-good-merge} depicts another example of a successful merge of states.

At the same time, the algorithm detected state changes even when most of the content had not been changed. For instance, a state change occurs when the Fandango application is provided with information on the user's location. Prior to this, the user is asked about her location via a form. When this form is filled and sent, the application state is changed, and geotargeted information is presented instead. This may affect the security vulnerabilities in a page; hence the need to consider these two states as different. 

\begin{figure}
\centering
\includegraphics[height=6.9cm,keepaspectratio]
{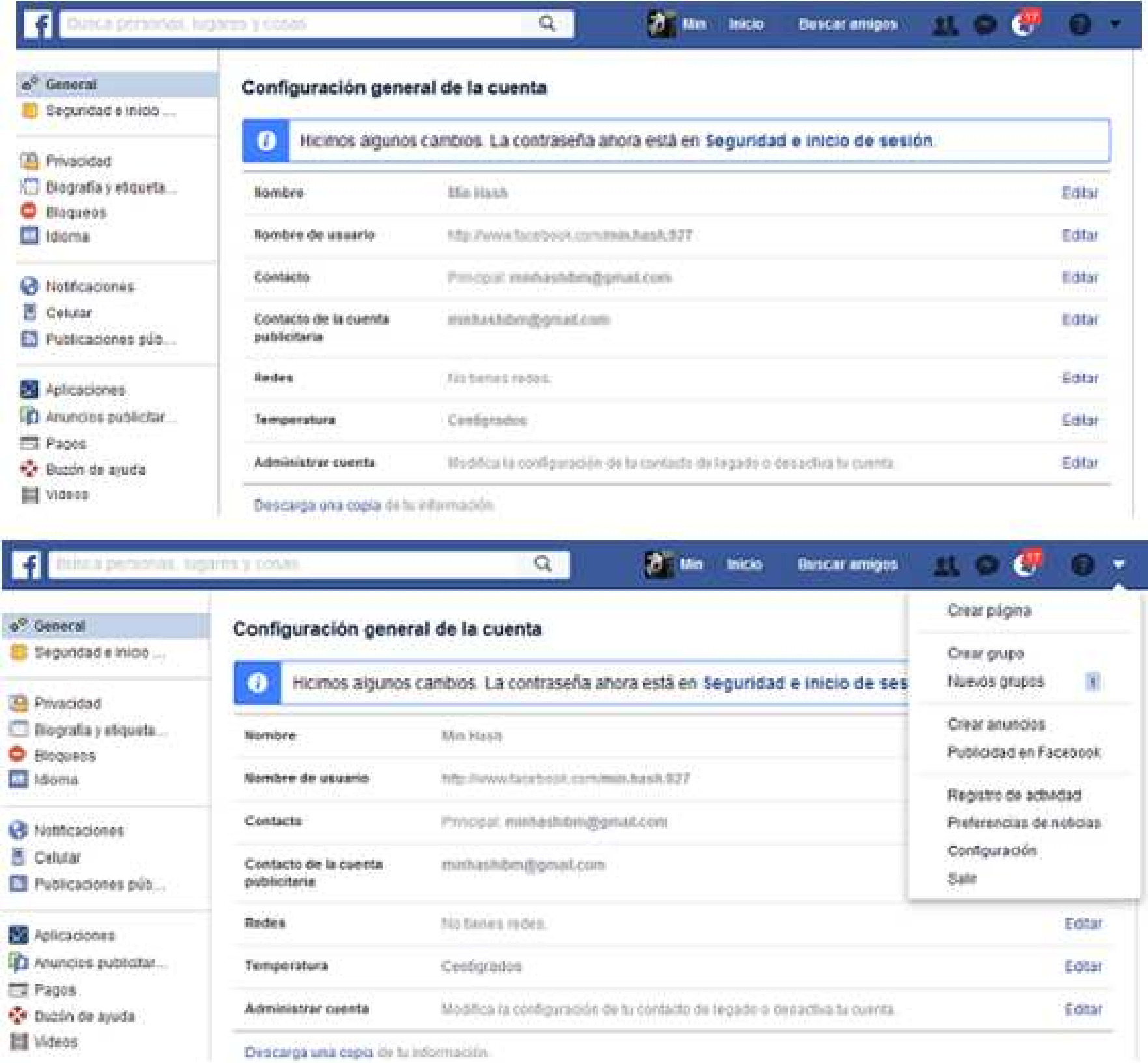}
\caption{Successful split of two nearly similar states using the MinHash algorithm. (a) Upper part: account setting view of a Facebook user. (b) Lower part: the same view with an additional menu open, adding a new functionality to the previous state. }
\label{FB-good-split}
\end{figure}

A successful split can also be found in the results of the Facebook scan. The privacy settings state is altered if an additional menu is open, and this change was detected by our method (see Figure \ref{FB-good-split}). Such a case frequently occurs in complex online applications, where there are several actions that are common to many views, e.g., opening menus or enabling a chat sidebar. Therefore, the number of possible states can be high. However, almost every combination differs from the others, so they are usually considered different states. We discuss a method to safely reduce the number of reported states in section \ref{conclusions}. 

There were also cases where the MinHash algorithm incorrectly split or merged states. One such case was during the Facebook scan, when analyzing two states that are accessed through the security settings page. One state allows choosing trusted friends, and the other enables a password change. Despite their similarity, there are different actions that can be done in each of these states. Hence, they should be counted as two different states, and were mistakenly merged into one. In the opposite direction, some states were split during the scan into multiple groups because of insignificant changes between different responses. Section \ref{conclusions} suggests several explanations for these incorrect results.

\subsubsection{Results of Other Hash-Based Approaches}
\noindent As was pointed in \cite{benjamin2010some}, when the equivalence relation is too stringent, the scan might suffer from a state explosion scenario, which requires a significant amount of time and memory. This observation accurately describes the scans involving the other two hashing methods, which often resulted in a state explosion scenario. Even when they included most of the states reported by the MinHash scans, they also contained hundreds of redundant states. This is due to the fact that the crawler exhaustively enumerated variations of a very limited number of states. The Facebook and Fandango scans reported hundreds of states, with more than 95\% of the states being duplicate content. Moreover, since the Facebook application is very complex and diverse, the state explosion scenario caused these scans to terminate due to memory problems.

It is interesting to analyze the performance of the na\"ive approach and the DOM Uniqueness algorithm when crawling the list of starred (tracked) projects of GitHub users. While it is clear that the na\"ive approach failed to merge two views with different number of starred projects, it is surprising to see that the DOM uniqueness algorithm also did not completely succeed in doing so. The reason for that is because the items of the list do not necessarily have the same DOM structure. Therefore, there is not always a common repeating pattern between different lists, and the lists were not always reduced to the same DOM canonical form. 

\subsubsection{Results of Non Hash-Based Tools}
\noindent The efficiency of the non hash-based scanners dropped significantly comparing to their performance on simple RIAs. This is due to the fact that complex online applications pose challenges that are still not well-addressed in these crawlers. CRAWLJAX requires manual configuration of tailored comparators to perform better on certain applications. Moreover, CRAWLJAX is less scalable due to its all-pairs comparison approach. 

The efficiency rates of the j{\"A}k scans were not high, as the assumption that similar states must share the same normalized URL is not always true. In the Fandango application, for example, pages providing movie overviews contain the movie name as a path parameter, and thus are not identified as similar. The same problem also occurred when j{\"A}k scanned Netflix. Multiple pages of this website offer the same content and the same set of possible user interactions, albeit they are written in different languages. 

The crawling components of j{\"A}k and CRAWLJAX could not login to the Facebook application properly, so a fair comparison is not possible in that case.

\section{\uppercase{Conclusions}}
\label{conclusions}

\noindent In this paper we present a locality-sensitive hashing scheme for detecting duplicate content in the domain of web application security scanning. The method is based on MinHash sketches that are stored in a way that allows an efficient and robust duplicate content detection. 

The method is theoretically proven to be less stringent than existing DOM state equivalence strategies, and can therefore outperform them. This was also empirically verified in a series of experimental scans, in which other methods, whether hash-based or not, either completely failed to scan real-world applications, or constructed large models that did not capture their structure. As opposed to that, the MinHash scheme enabled successful and efficient scans.  Being able to better detect similar content prevents the MinHash algorithm from exploring the same application state over and over. This way, the scans constantly revealed new content, instead of exploring more views of the same state. The MinHash scans got beyond the first layers of the application, and did not consume all the scan time on the first few states encountered.

Reducing the scan time and the complexity of the constructed application model does not always come without a price. The cost of using an LSH approach might be an increase in the number of different states being merged into one, and this could lead to an incomplete coverage of the application. However, the risk of a more strict equivalence strategy is to spend both time and memory resources on duplicate content, and thus to achieve poor coverage. 

This risk can be mitigated by better optimizing parameter values. The value of $k$, the shingle length, can be optimized to the scanned application. One can take factors such as the average length of a DOM state, or the variance in the number of different elements per page, when setting the value of $k$ for a given scan. Tuning the similarity threshold per application may decrease the number of errors as well. Of course, the probabilistic nature of the method also accounts for some of the incorrect results, as the Jaccard similarity estimations are not always accurate. Increasing the number of hash functions can reduce this inaccuracy.

The LSH scheme can be applied to any web application. However, there are applications in which its impact is more substantial. Black-box security scanners are likely to face difficulties when scanning complex applications. Such applications often heavily rely on technologies such as AJAX, offer dozens of possible user interactions per page, and contain a great amount of near-duplicate content. The MinHash algorithm can dramatically improve the quality of these scans, as was the case with Facebook or Fandango. For applications with less variance in the structure of pages that belong to the same application state, such as Netflix, the DOM uniqueness state equivalence mechanism can still perform reasonably. But even in these cases, the MinHash algorithm reaches the same coverage more efficiently. 

A locality-sensitive hashing function helps in a better detection of similarity between states. However, if a state is actually a container of a number of components, each having its own sub-state, then the number of states in the application can grow exponentially. It seems reasonable that applying our algorithm on every component separately would give better results. In this context, the component-based crawling \cite{moosavi2014indexing} could be a solution for decomposing a state into several building blocks, which will all be analyzed using the MinHash algorithm.

The tested implementation of the MinHash scheme is very basic. There is a need to take more information into consideration when constructing the DOM string representation of a response. For example, one can mark part of the elements as more relevant for security oriented scans, or use the value of the attributes as well. Theoretically, two pages can have a very similar DOM and differ only in the event handlers that are associated with their elements. Although this usually do not occur in complex applications, there were some rare cases where different states were mistakenly merged into one state due to that reason. Another potential improvement is to detect which parts of the DOM response are more relevant than others. By doing so we can mark changes in elements that are part of a navigation bar as less significant than those occurring in a more central part of the application.

We believe that incorporating these optimizations in the MinHash scheme would make this approach even more robust and accurate. Combined with more sophisticated crawling methods such as component-based crawling, security scans can be further improved and provide useful information for their users.

\section*{\uppercase{Funding}}
\noindent This study was supported by fellowships from the Edmond J. Safra Center for Bioinformatics at Tel-Aviv University, Blavatnik Research Fund, Blavatnik Interdisciplinary Cyber Research Center at Tel-Aviv University, and IBM.

\section*{\uppercase{Acknowledgements}}

\noindent We wish to thank Benny Chor and Benny Pinkas for their helpful suggestions and contribution. We are grateful to Lior Margalit for his technical support in performing the security scans. The work of Ilan Ben-Bassat is part of Ph.D. thesis research conducted at Tel-Aviv University.

\bibliographystyle{apalike}
{\small
\bibliography{minhash}}

\end{document}